# X-ray projection imaging of metal oxide particles inside gingival tissues

Jarrod N. Cortez,[1] Ignacio O. Romero,[2] Md Sayed Tanveer,[5] Chuang Niu,[5] Cássio Luiz Coutinho Almeida-da-Silva,[3] Leticia Ferreira Cabido,[4] David M. Ojcius,[3] Wei-Chun Chin,[2] Ge Wang,[5] Changqing Li[1,2*]

[1] *Quantitative and Systems Biology, University of California, Merced, Merced, CA.*
[2] *Department of Bioengineering, University of California, Merced, Merced, CA.*
[3] *Department of Biomedical Sciences, University of the Pacific, San Francisco, CA.*
[4] *Department of Oral and Maxillofacial Surgery, University of the Pacific, San Francisco, CA.*
[5] *Department of Biomedical Engineering, Biomedical Imaging Center, Center for Biotechnology and Interdisciplinary Studies, Rensselaer Polytechnic Institute, Troy, NY 12180, USA.*

*Corresponding author: cli32@ucmerced.edu

**Abstract**

There is increasing recognition that oral health affects overall health and systemic diseases. Nonetheless it remains challenging to rapidly screen patient biopsies for signs of inflammation or the pathogens or foreign materials that elicit the immune response. This is especially true in conditions such as foreign body gingivitis (FBG), where the foreign particles are often difficult to detect. Our long term goal is to establish a method to determine if the inflammation of the gingival tissue is due to the presence of a metal oxide, with emphasis on elements that were previously reported in FBG biopsies, such as silicon dioxide, silica, and titanium dioxide whose persistent presence can be carcinogenic. In this paper, we proposed to use multiple energy X-ray projection imaging to detect and to differentiate different metal oxide particles embedded inside gingival tissues. To simulate the performance of the imaging system, we have used GATE simulation software to mimic the proposed system and to obtain images with different systematic parameters. The simulated parameters include the X-ray tube anode metal, the X-ray spectra bandwidth, the X-ray focal spot size, the X-ray photon number, and the X-ray detector pixel. We have also applied the de-noising algorithm to obtain better Contrast-to-noise ratio (CNR). Our results indicate that it is feasible to detect metal particles as small as 0.5 micrometer in diameter when we use a Chromium anode target with an energy bandwidth of 5 keV, an X-ray photon number of $10^8$, and an X-ray detector with a pixel size of 0.5 micrometer and 100 by 100 pixels. We have also found that different metal particles could be differentiated from the CNR at four different X-ray anodes and spectra. These encouraging initial results will guide our future imaging system design.

## 1. Introduction

Oral health is overlooked far too often when compared to maintaining one's overall health, despite evidence that it affects non-oral systemic health. Therefore, dentistry requires better detection methods to ensure that any complications are properly diagnosed at early states of oral disease [1,2]. This study aims to develop a new method for imaging the gingival tissue to detect and differentiate foreign metal particles inside gingival tissues, which can trigger an inflammatory immune responses seen in conditions such as gingivitis.

In the past, different research groups have sought to detect metal particles in gingivals tissues from dental implants made of metals or alloys such as titanium (Ti) [3-5]. In 2003, Olmedo et. al. investigated how the presence of Ti in macrophages near implant sites in the soft tissue reflected the immune response of the body to the presence of a foreign material [3]. They and other scholars have also investigated how the wear of these implants was causing the immune system of patients to react to the foreign debris in different parts of the body [3,4]. The foreign debris was characterized through methods such as Energy Dispersive X-ray (EDX) which requires the samples to be coated by a metal such as gold [3]. In 2010, Flatebø et. al. utilized laser ablation inductively coupled plasma mass spectroscopy (LA-ICP-MS) to identify the elemental composition of samples in a way that the metal particles inside samples could be differentiated. They also employed high-resolution optical darkfield microscope (HR-ODM) to image samples as well as Scanning Electron Microscopy (SEM) to measure the particle sizes [4]. In 2016, Fretwurst et. al. studied the effects of peri-implantitis, a condition caused by implants in a patient, by analyzing the metal elements in patient bone and mucosal tissues near the implant. They performed their analysis with Synchrotron radiation X-ray fluorescence microscopy (SRXRF) and Polarized Light Microscopy (PLM). Samples for their study were obtained during surgery as all the patients included were already confirmed to have severe peri-implant diseases [5].

Although these studies have linked metal oxides to immune responses, the equipment they have used are very specialized, expensive, time-consuming, and not available for many scholars and clinics. Some methods such as LA-ICP-MS needs special treatment of the samples, preventing analysis with other methods. In this paper, we proposed a multiple energy x-ray projection imaging system to address these challenges. The proposed system is portable, affordable, and capable of particle differentiations.

During foreign body gingivitis (FBG), gingival tissues become inflamed due to the immune response to foreign metal oxides such as silica which are known carcinogens [2,6-8]. However, there is no established method for clinicians to diagnose rapidly whether this specific form of gingivitis is caused by an infection or a foreign material, a shortcoming that this study will address. Furthermore, the lesions from gingivitis have several phases, with the final one being a transition from gingivitis to the more severe periodontitis which leads to a loss of attachment and bone destruction [9]. The review by Ivanovski covers the effects of bone loss as a result of dental implants, peri-implantitis, with a detailed description of the body's immune response to these implants as well as other foreign materials [10]. The proposed X-ray projection imaging shares the same field of view of the fluorescence optical microscope so that we can study the immune responses and their correlations with the foreign metal particles inside samples. The system performance is evaluated with numerical simulations in this paper.

## 2. Methods

### 2.1 GATE software

The GEANT4 Application for Tomographic Emission (GATE) was used in this study to simulate the attenuation-based X-ray projection imaging [11]. The GATE package was downloaded from opengatecollaboration.org and was run in our cluster with 20 cores and 64 GB memory. We used macro files to program the systematic parameters like X-ray energy and photon number and positions. In our studies, we have assumed an ideal detector which stopped all arrival X-ray photons and recorded their energy and positions for future processing.

The simulation follows the standard components needed for a CT simulation, the imaging object, X-ray source, detector, tissue sample including the metal oxides and the polystyrene substrate sheet. Each simple geometry can be created in GATE and then the material of any given geometry can be assigned if the elemental composition is known. More details about GATE simulation setup and objects are described below.

### 2.2 X-ray imaging system in GATE

To optimize the system parameters used as guidance for our future experimental system design, we simulated the following system parameters by studying how these parameters affect the image quality: the target anode used to produce X-rays, the X-ray tube focal spot size, the spectra for each of these targets, the number of X-rays initialized, and the size of the pixels for the detector.

### 2.2.1 X-ray imaging system in GATE

As depicted in Fig. 1(top), the system makes use of a cone-beam X-ray source with a very small half-cone angle of 0.11 degrees with the source 10 mm ahead of the origin of the simulation. The focal spot size of the source is also set to be 4 micrometers. The sample patch which contains the metal oxides is set at the origin of the instance. The polystyrene sheet is centered 87.5 microns behind the sample sheet just to prevent the geometries from overlapping since the sheet is 170 micrometers thick. The detector is then 1 mm behind the sample.

The substrate sheet is added to ensure that the projection can be acquired with a surface for the sample to lie upon. After testing materials such as carbon and glass, it was determined that the low attenuation of polystyrene would be the best for a substrate that the sample could be on during projection acquisition. The sheet is 10 mm by 10 mm in length and width with a thickness of 0.17 mm. The tissue sample mimics soft tissue and is made of water. The sample is 2 mm by 2 mm with a thickness of 5 micrometers. Each of the metal oxides are discs 2 micron in diameter with a thickness of 2 micrometers. We have selected five metal oxides which were found most likely in many dental products. The five metal oxides are Silicon dioxide, Aluminum oxide, Titanium dioxide, Zinc oxide, and Iron oxide. Their positions in the sample are indicated in Fig. 1(bottom, left) and kept consistent for each simulation except for when the limits of detection simulations were performed.

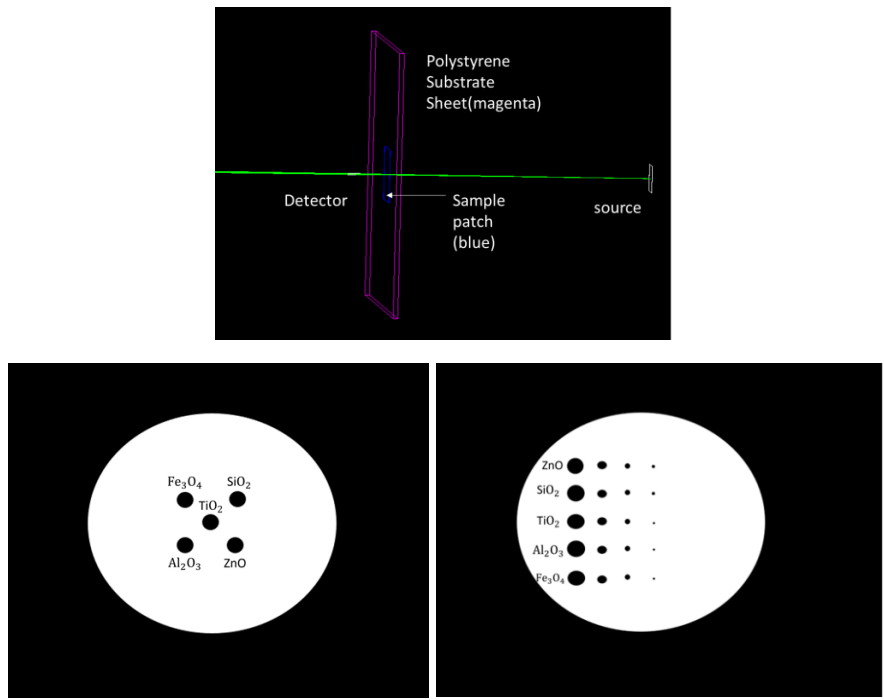

*Fig. 1. (Top) GATE simulation setup for numerical simulation studies. Cone beam X-ray tube and ideal X-ray detector were used. Focal spot size and detector pixel size and their effects have also been studied. (Bottom, left) The ground truth image for the Metal oxide particles in the simulation for 2 micron diameter metal oxides. (Bottom, right) The ground truth image for particle size detection. The first column includes samples with a 2 micron diameter, the second column is 1 micron diameter, third is 0.5 micron and the last is 0.25 micron in diameter. The x-ray source had a 4 micrometer focal spot size.*

To test the smallest single particle the X-ray projection imaging can detect, we perform the X-ray projection imaging of particle targes with different sizes from 0.25 to 2 micrometers in diameters as shown in Fig. 1 (bottom, right). The leftmost column particles have diameter of 2 micrometers. The 2$^{nd}$ leftmost column targets have a diameter of 1 micrometer. The 3$^{rd}$ leftmost column targets have a diameter of 0.5 micrometer. And the rightmost column targets have a diameter of 0.25 micrometers. As shown in Fig. 11, the same metal particles are placed in one row with five total metal particles from top row to bottom row. The distances between rows and columns of these metal particle targets are 5 micrometers. The parameters for the simulation were as follows, the 5 kV bandwidth Cr source, 100 by 100 detector with 0.5 micron detector pixel length, 4 micron focal spot size, and various x-ray photon numbers as $10^9$, $10^8$, $10^7$, and $10^6$.

*2.2.2 X-ray spectra*

Fig. 2 plots the full energy spectra of X-ray tube with different four anode target metals: Chromium (Cr) (top left), Copper (Cu) (Top right), Gold (Au) (Bottom left), and Molybdenum (Mo) (Bottom right). The horizontal axis indicates the X-ray photon energy in units of keV. The vertical axis indicates the normalized photon number per energy bin. We have also simulated narrowed X-ray spectra cases with energy bandwidths of 5 keV and 10 keV, respectively, for all four target metals. For Cr and Cu, the narrowed energy band is centralized at the highest energy peak. For Au, the narrowed energy band is centralized at the peak of 11.5 keV. For Mo, the narrowed energy band starts from 0 keV to include the peak of 2.3 keV.

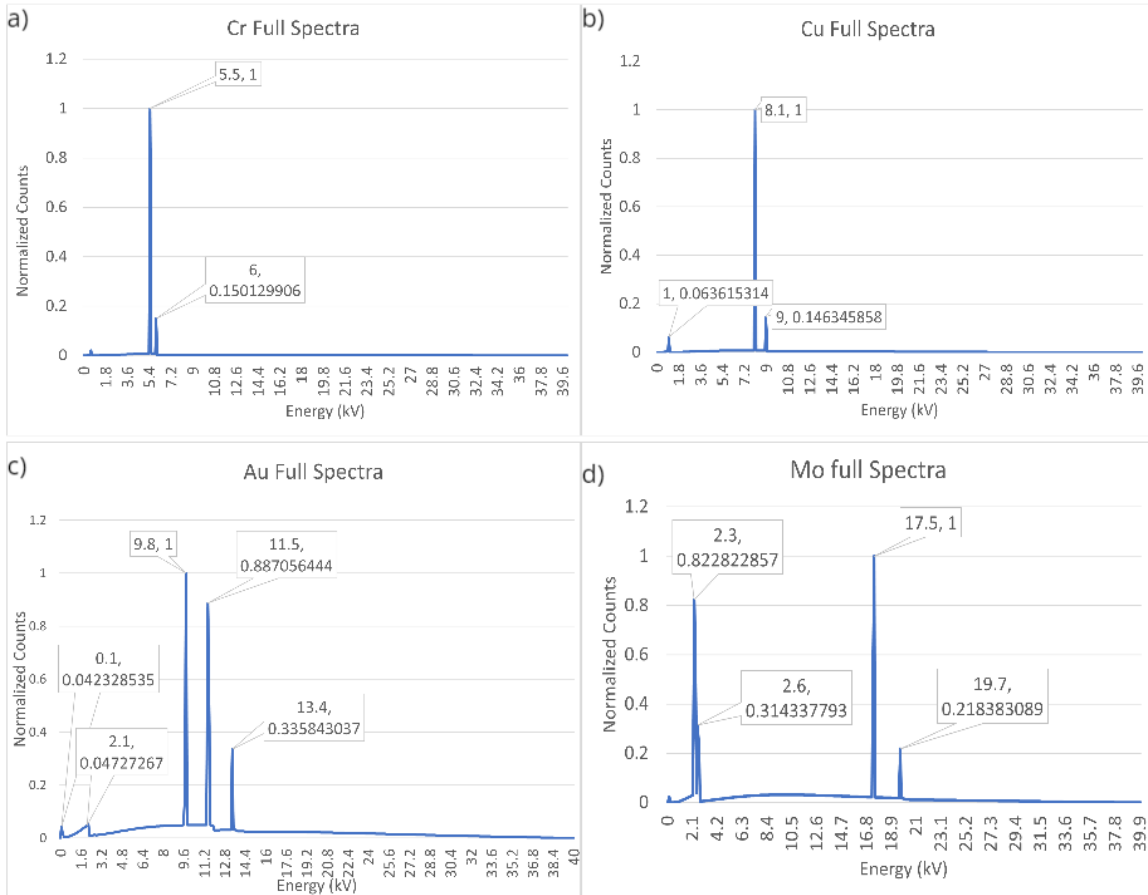

Fig. 2. X-ray spectra for each of 4 target metals: Cr (top left), Cu (top right), Au (bottom left), and Mo (bottom right).

### 2.2.3 X-ray detector

The X-ray detector is small with a varying number of pixels which can be adjusted after the simulation is run by editing the MATLAB script we used for reconstruction. This detector is made of CZT with a thickness of 1 mm. The X-ray interaction position in detector of each photon is recorded. We record the positions with the Phase Space Actor in GATE which allows us to use MATLAB script to adjust the number of pixels in the X-ray projection imaging because the detector does not have a predetermined number of pixels.

### 2.3 Attenuation-based X-ray imaging

To investigate how different system parameters affect the image quality and the differentiation of different particles inside tissues, we have changed the following parameters: the target anode used to produce X-rays, the spectral width for these targets, the X-ray tube focal spot size, the number of X-rays initialized, and the pixel size of the detector.

We have investigated four different target metals of X-ray tube thus four different X-ray spectra, and changed three different spectra bandwidths for each of the spectra, as described in section 2.2.2. We have also studied three different X-ray focal spot sizes such as 1, 2, and 4 micrometers in diameter. For each case, the X-ray photon number was changed from $10^6$ to $10^9$ with a step size of 10 times. As described in section 2.2.3, the Phase Space Actor in GATE allows us to change the detector's pixel size after we made the measurements. We have selected the detector sizes as 0.25, 0.5, 1, and 2 micrometers.

*2.4 Denoising algorithm*

Using denoising algorithms, the noise in the image can be differentiated from the oxides we wish to identify. This requires training the algorithms with data previously acquired during the simulations so that key traits can be classified by the algorithm. The result is an image with a higher contrast-to-noise ratio (CNR) as the oxides are separated from the noise producing a higher quality image.

Here, our goal for denoising is not only to increase CNR for the metal oxides but also for them to have enough distinction to be recognized separately. For this reason, we apply four different denoising techniques, specifically: 2D wavelet denoising (biorthogonal wavelet), Denoising Convolutional Neural Network (DnCNN) by Zhang et al (MATLAB implementation) [12], Block-matching and 3D filtering (BM3D) by Dabov et al [13] and Noise2Sim by Niu et al [14].

However, for this paper, only the best results are shown. Both DnCNN and BM3D tend to create a more "visually appealing" denoised image, but they often fail to retain the quantitative properties inside the projection images that are necessary for classifying different metal oxides. In this regard, Noise2Sim provides not only an increase in the mean CNR but also a relative reduction in their variances for all photon numbers, which allows the most separation in their values. Thus, Noise2Sim is the most suitable denoising algorithm for our purpose.

*2.5 Evaluation criteria: contrast-to-noise ratio*

To evaluate the image quality, we calculate the CNR ratio for both attenuation X-ray images and phase-contrast X-ray images. Following a paper in 2016 [15], we used the following equation:

$$CNR = \frac{Mean(x_{ROI}) - Mean(x_{ROB})}{\sqrt{\omega_{ROI} Var(x_{ROI}) + (1 - \omega_{ROI}) Var(x_{ROB})}} \quad (1)$$

where ROI and ROB indicate region of interest and region of background, respectively. The weight factor ω was calculated by the fraction of ROI related to the ROB. Typically, we will use a region of pixels dependent on the number of pixels that are inscribed within a sample which differs based on how many pixels are in the image for CNR calculations. All the calculations were performed by MATLAB.

## 3. Results

We have investigated how the different systematic parameters affect the CNR of five different metal oxides, numerically. We describe our results in the following subsections. Each simulation keeps the five metal oxides in the same arrangement: iron oxide at the top left, silicon dioxide in the top right, titanium dioxide in the center, aluminum oxide in the bottom left and zinc oxide in the bottom right. This way each future simulation can be easily compared to previous ones to examine the effects of source spectra, focal spot size, and other parameters on the projection quality.

Each of the attenuation-based projections produced over the course of the study were created through MATLAB. The code creates two images an attenuation-image and a projection image with normalized intensities that allows each projection to be compared using the same scale.

### 3.1 Effects of X-ray anode metals

Fig. 3 depicts the X-ray projection images of the samples using $10^8$ X-ray photons and 4 micrometers focal spot size and 5 keV energy bandwidth for the X-ray anode metal of Chromium (a), Molybdenum (b), Copper (c), and Gold (d), respectively. The X-ray detector for this study has a pixel size of 0.5 micrometers and 100 by 100 pixels. As expected, we see the differences among CNRs for different particles due to the different energy peaks and different energy ranges for four different X-ray anode metals, which is very helpful for differentiation between the different particles. The calculated CNRs are listed in Table 1, from which we see that particles with light atomic number have larger attenuations/CNR in the low energy band and particles with heavy atomic number have higher attenuation/CNR in higher energy band.

**Table 1. The CNR values for each metal oxide examined in a 100 by 100 detector pixel reconstruction**

| Source<br>Metal | Mo<br>(Standard Deviation) | Cu | Cr | Au |
|---|---|---|---|---|
| Al | 4.3648 | 7.0448 | 6.7847 | 6.7565 |
|  | (0.4902) | (0.6416) | (0.6885) | (0.5270) |
| Fe | 10.2891 | 17.1275 | 18.2211 | 17.8851 |
|  | (1.0154) | (0.8650) | (1.4855) | (1.7107) |
| Si | 5.7683 | 8.8874 | 8.7102 | 8.6084 |
|  | (0.6822) | (0.7281) | (0.6327) | (0.9212) |
| Ti | 9.4123 | 16.6091 | 27.3843 | 19.0917 |
|  | (1.2162) | (0.9254) | (2.0064) | (1.8566) |
| Zn | 8.4891 | 13.5311 | 10.7164 | 15.0830 |
|  | (0.6582) | (0.9520) | (0.4379) | (1.4182) |

**$10^8$ X-ray photons, 4 micrometers focal spot size, and 5 keV energy bandwidth were used for all. The source anode is indicated by the first column and each column after is the CNR of that oxide when each source was used.**

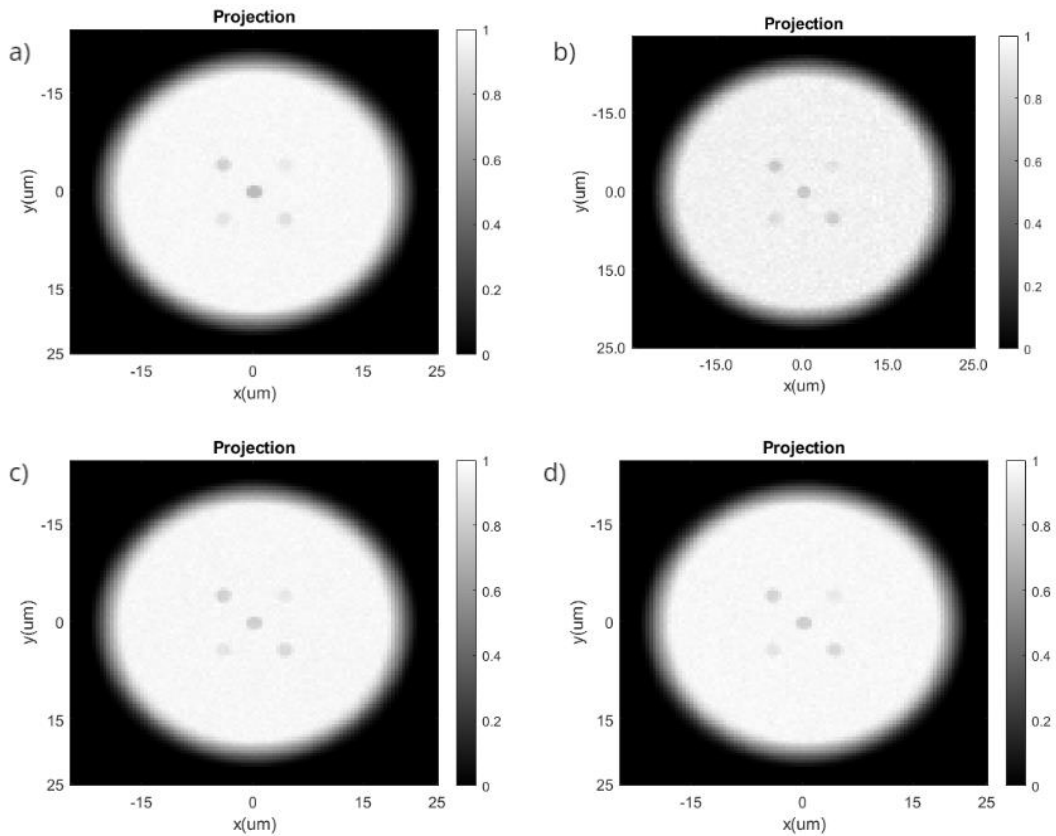

Fig. 3. Attenuation based Projection of the metal oxide samples using 5 kV energy band spectra produced by the (a). Chromium, Cr, target anode, (b). Molybdenum, Mo, target anode, (c). Copper, Cu, target anode, and the (d). Gold, Au, target anode from the Sigray Source using $10^8$ X-rays. Targets kept in the same arrangement as Figure 3.

## 3.2 Effects of X-ray spectra

The effects of the different X-ray spectra on contrast are shown in Fig. 4 where different energies of the same X-ray source were used to investigate whether this affected the CNR of the metal particles. For the Chromium X-ray source, the low attenuation targets such as Aluminum and Silicon do not have good contrast when the full spectra (40 keV) is used. When the energy spectra is truncated from 0 to 10 keV, these two low contrast targets appear more clearly. We have also observed that the contrast becomes best when a 5 keV spectra from 2.5 keV to 7.5 keV is used. We have calculated all the CNR for 4 types of X-ray tubes and their corresponding three types of energy spectra and listed the results in Table 2. Our results also reveals that there are certain energies within the overall spectra which allow some metal oxides to have an increase in their CNR while others are lower. For example, for Al and Fe of the 10 keV and 5 keV cases, there is an increase in the CNR to 19.5 for Fe at 10 keV, while at 5 keV it drops to 13.0. By contrast, the CNR of Al at 10 keV is 5.6 while it is 6.9 for the 5 keV case. All these results indicate that the energy band of the X-ray source anode has significant impacts on the CNR of metal oxides.

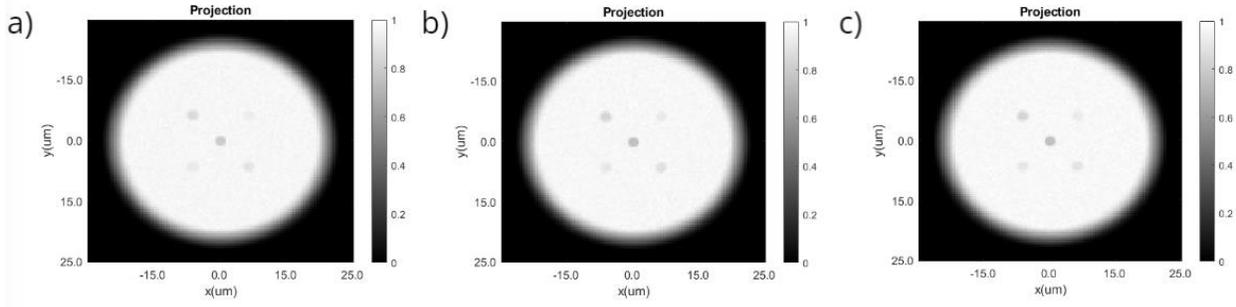

Fig. 4. Attenuation based Projection of the metal oxide sample using (a) the full 40 kVp spectra, (b) the 10 kV energy band spectra produced, and (c) the 5 kV energy band for the Cr target anode with $10^8$ X-rays, 4-micron focal spot size.

**Table 1. The CNR values of the five metal oxides**

| Energy of Cr source | CNR of Al (Standard Deviation) | CNR of Fe | CNR of Si | CNR of Ti | CNR of Zn |
|---|---|---|---|---|---|
| 40 keV | 4.240 (0.6891) | 13.5108 (1.8120) | 6.1807 (0.6630) | 17.5173 (2.3201) | 8.3160 (1.2813) |
| 10 keV | 5.1265 (0.2471) | 14.4045 (0.6899) | 6.6470 (0.4626) | 20.9338 (0.7918) | 9.1214 (0.4007) |
| 5 keV | 6.7847 (0.6885) | 18.2211 (1.4855) | 8.7102 (0.6327) | 27.3843 (2.0064) | 10.7164 (0.4379) |

**The energy of the Cr source spectra was well set to full then truncated to 10 keV and a 5 keV bandwidth. Based on a 100 by 100 pixel detector, 0.5 micrometer pixel size, $10^8$ X-ray photons, and 4 micrometer focal spot size were used for all.**

### 3.3 Effects of X-ray photon number

To investigate how the X-ray photon number in each projection affects the target detection, we have increased the X-ray photon number from $1.0 \times 10^6$ to $1.0 \times 10^9$ with a multiple factor of 10 for each X-ray anodes and energy spectra. In this simulation, we used the same X-ray focal spot size of 4 μm and the same X-ray detector pixel size of 0.5 μm. The results of a typical example for 5 keV, Cr X-ray source with four different X-ray photon numbers are depicted in Fig. 7. Their CNRs of the five metal targets calculated from Fig. 5 are listed in Table 3. For the target of Fe at $10^9$ x-rays, the CNR is 47.6 whereas with $10^8$ x-rays the contrast ratio is only 13.0. For the target of Al on the other hand, the CNR is 5.9 at $10^9$ x-rays then has a higher CNR for the case of $10^8$ x-rays which may be due to the overall increase in X-ray photon number for each pixel in the projection since the detector dimensions remained constant throughout these simulations. Furthermore, as expected a lower X-ray count causes a decrease in most CNR, though not to the same degree as the targets of Ti has a high contrast at the X-ray photon number of $10^7$.

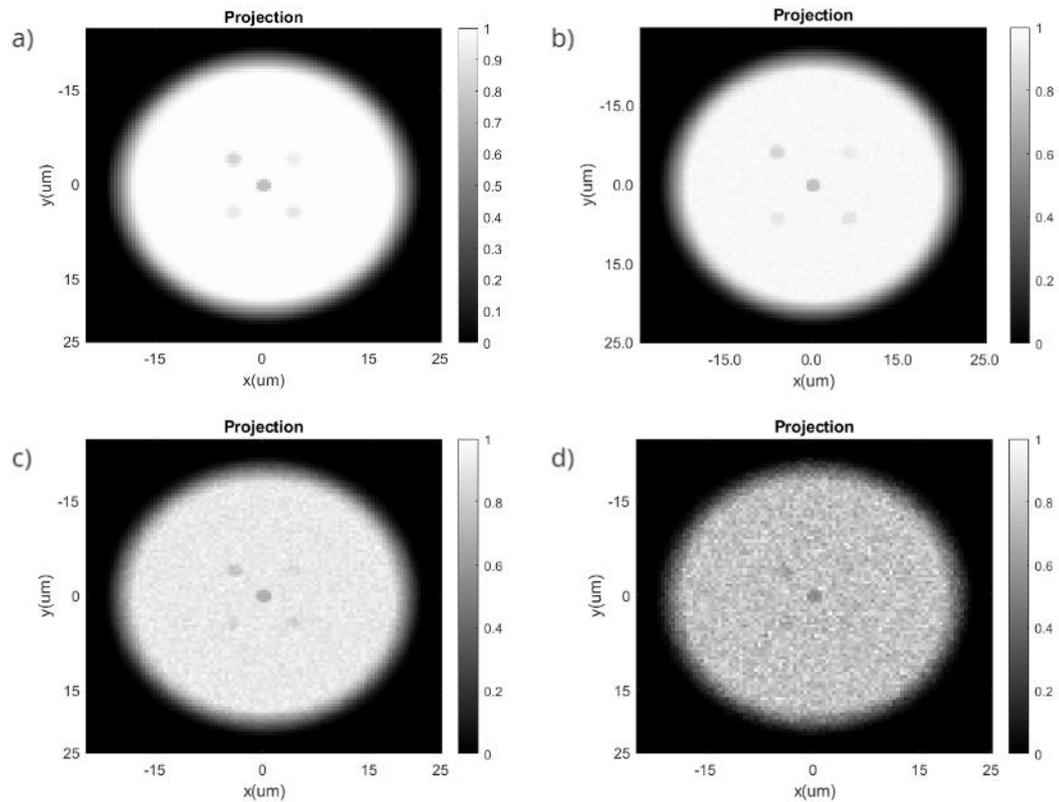

Fig. 5. Attenuation based Projection of the metal oxide sample using a 5 kV energy band spectra produced by the Cr target anode from the Sigray Source using (a). $10^9$, (b) $10^8$, (c). $10^7$, and (d). $10^6$ X-Rays.

**Table 3. All calculated CNR for the metal oxides when the Cr source when the x-ray photon number was altered**

| X-ray Count | CNR of Al (Standard deviation) | CNR of Fe | CNR of Si | CNR of Ti | CNR of Zn |
|---|---|---|---|---|---|
| $10^9$ | 20.9694 (0.3322) | 58.0322 (0.2189) | 37.7972 (0.2189) | 88.1443 (3.3499) | 33.9443 (0.3150) |
| $10^8$ | 6.7847 (0.6885) | 18.2211 (1.4855) | 8.7102 (0.6327) | 27.3843 (2.0064) | 10.7164 (0.4379) |
| $10^7$ | 1.8870 (0.7847) | 6.4686 (1.0128) | 2.8590 (0.6133) | 8.9391 (1.6853) | 3.4604 (0.5891) |
| $10^6$ | 0.7150 (0.3173) | 1.8595 (0.3017) | 3.1382 (0.5076) | 3.9342 (0.6176) | 11.4496 (0.0810) |

**Based on a 100 by 100 pixel detector, 0.5 micrometer pixel size, 4 micrometer focal spot size, and 5 keV energy bandwidth were used for all**

### 3.4 Effects of X-ray focal spot size

To investigate how the X-ray tube focal spot size affect the CNR, we have reduced the X-ray focal spot size from 4 μm to 0.25 μm with a multiple factor of 0.5. In this study, we used the Cr source anode with the 5 keV energy bandwidth, the detector with a pixel size of 0.5 μm and 100 by 100 pixels, and the x-ray photon number of $10^8$. The X-ray projection images for the cases with five different focal spot sizes are depicted in Fig. 8 and the corresponding CNR for each of five targets from these five images are calculated and listed in Table 4. From Fig. 6, we see that the edges of the projection appear less blurry when using a finer X-ray cone beam since the beam becomes smaller which allows for sharper details within the projection. Once again, the CNR of each metals change as shown in Table 4. From this table, we see that the overall reduction of CNR by increasing focal spot size from 0.25 to 4 μm is not significant, which means that it is acceptable to use a large focal spot size of 4 μm in the future X-ray imaging system design.

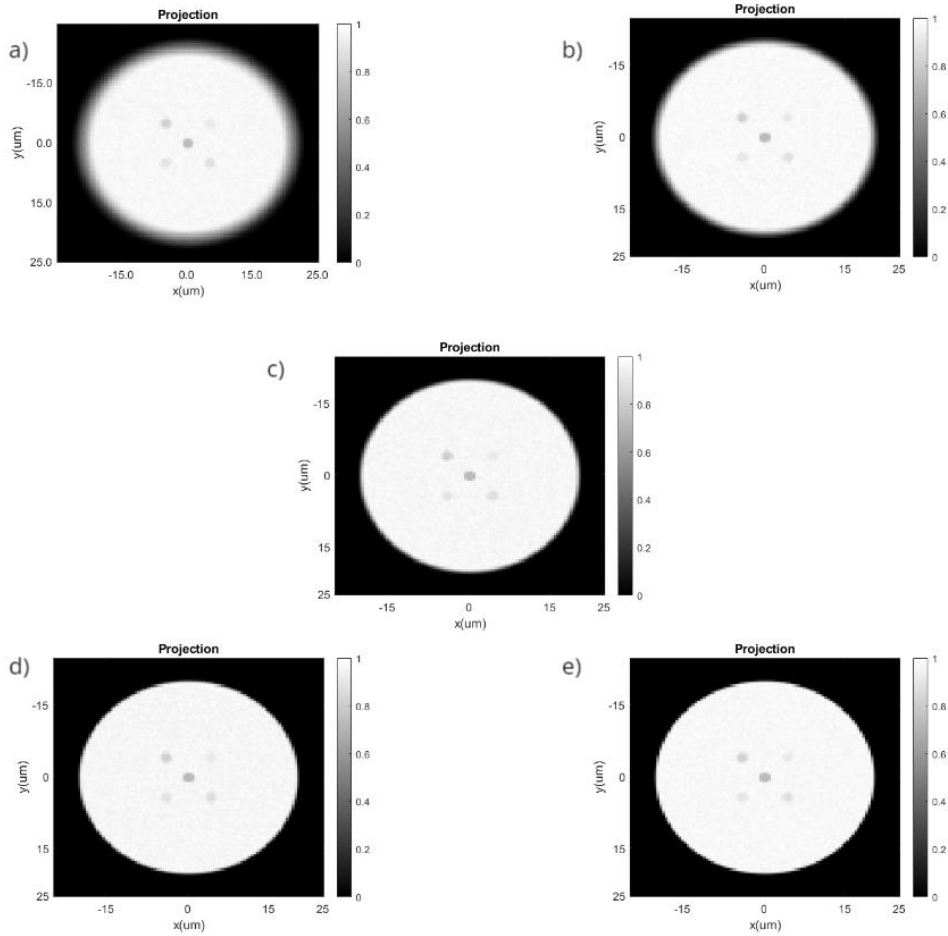

Fig. 6. Attenuation images produced in MATLAB of the 5 metal particles of interest with 5 kV truncated spectra for each target anode with an a) 4 micron focal spot size , b) 2 micron focal spot size, c) 1 micron focal spot size, d) 0.5 micron focal spot size, and e) 0.25 micron focal spot size.

**Table 2. Calculated CNR of all 5 metal oxides for differing focal spot sizes using the Cr source anode Based on a 100 by 100 pixel detector, 0.5 micrometer pixel size, $10^8$ x-rays, and 5 keV energy bandwidth were used for all.**

| Focal Spot Size (micrometers) | CNR of Al | CNR of Fe | CNR of Si | CNR of Ti | CNR of Zn |
|---|---|---|---|---|---|
| 4 | 6.7847 | 18.2211 | 8.7102 | 27.3843 | 10.7164 |
|   | (0.6885) | (1.4855) | (0.6327) | (2.0064) | (0.4379) |
| 2 | 7.8366 | 20.9003 | 10.2795 | 31.4289 | 11.2339 |
|   | (1.7035) | (3.8462) | (1.9419) | (6.7793) | (2.0899) |
| 1 | 6.8702 | 18.3082 | 8.7102 | 27.3843 | 10.7164 |
|   | (0.4996) | (1.2873) | (0.7468) | (2.2146) | (0.7784) |
| 0.5 | 7.0461 | 18.6752 | 9.1608 | 28.6224 | 10.9054 |
|   | (0.7916) | (2.0931) | (0.8926) | (2.7893) | (1.2207) |
| 0.25 | 5.9886 | 17.1340 | 8.0419 | 25.9724 | 9.7479 |
|   | (0.5365) | (1.6520) | (0.4493) | (1.1429) | (0.3291) |

### *3.5 Effects of X-ray detector pixel size*

Another area of this study was the size of each detector pixel, and Fig. 7 illustrates how the detector pixel size significantly impacts the CNR. This is because a larger detector pixel receives more X-ray photons if the total X-ray photon number is not changed. In this study, we used the Cr X-ray source with an X-ray energy bandwidth of 5 keV and a focal spot size of 4 μm, and an X-ray photon number of $10^8$. The X-ray detector has 100 by 100 pixels. Table 5 lists the calculated CNR values from Fig. 7. The general trend is that each sample becomes more visible when the detector pixel size decreases.

However, this does not translate to a guaranteed increase in CNR as we look at Al with a 200 by 200 versus a 100 by 100 pixel detector. With more pixels the contrast is 2.7 whereas it is about 6.9 in the 100 by 100 pixel detector case. This is because with more pixels a different ROI and ROB must be used for the CNR calculation allowing the value to change as the mean intensities in the ROB increase.

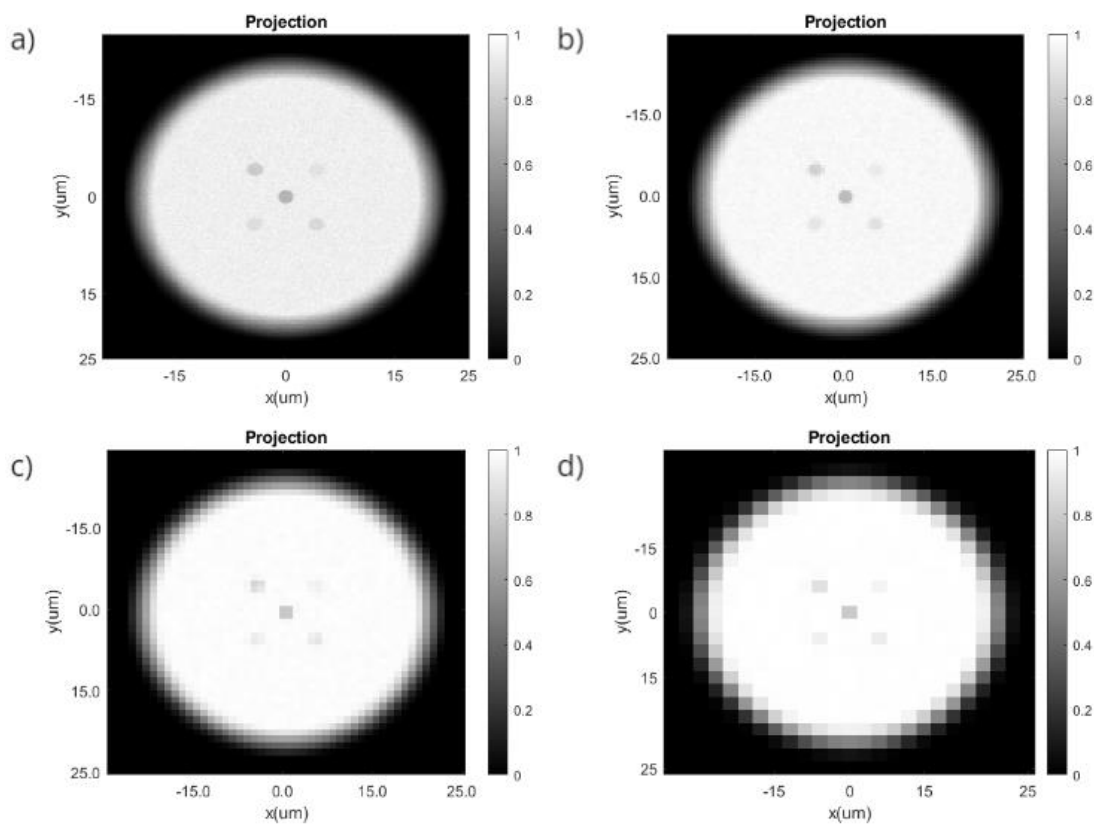

Fig. 7. Attenuation images produced in MATLAB for the 5 metal oxide particles of interest with 5 keV spectra for each target anode and with different detector pixel sizes: (a) 0.25 μm, (b) 0.5 μm, (c) 1.0 μm, and (d) 2 μm.

**Table 3. Metal oxide calculated CNR for different detector pixel sizes when using a Cr source**

| Detector pixel size (micrometers) | CNR of Al (Standard Deviation) | CNR of Fe | CNR of Si | CNR of Ti | CNR of Zn |
|---|---|---|---|---|---|
| 0.25 | 3.2089 (0.1209) | 9.1312 (0.3666) | 4.3645 (0.2675) | 13.7705 (0.6174) | 5.3767 (0.1887) |
| 0.5 | 6.7847 (0.6885) | 18.2211 (1.4855) | 8.7102 (0.6327) | 27.3843 (2.0064) | 10.7164 (0.4379) |

*4 micrometer focal spot size, 0.5 micrometer pixel size, $10^8$ x-rays, and 5 keV energy bandwidth were used for all.*

### 3.6 Differentiation of metal oxides from CNR plots

It was found that each spectra led to different projections with some particles showing more distinctly in the projection images with one target anode versus another shown clearly in the plots of Fig. 8. The contrast was determined to be the best when observing the projections constructed from the 5 kV energy band spectra which is a truncated spectra ranging from 2.5 to 7.5 kV. This produced some of the clearest images for the metal particles with the lowest attenuations due to their densities. The fact that each is different helps distinguish which one would work the best when attempting to identify unique particles.

To confidently identify a particle in a real sample there needs to be a way to separate it from the many other possible particles that may be in the sample. This led to the use of the CNR ratio across different parameters to ensure that the behavior of these metals was unique to each metal. The work done calculating the CNR did in fact show that the metals have their own unique CNR values with different energy spectra, and this changed when investigating different pixel sizes of the X-ray detector.

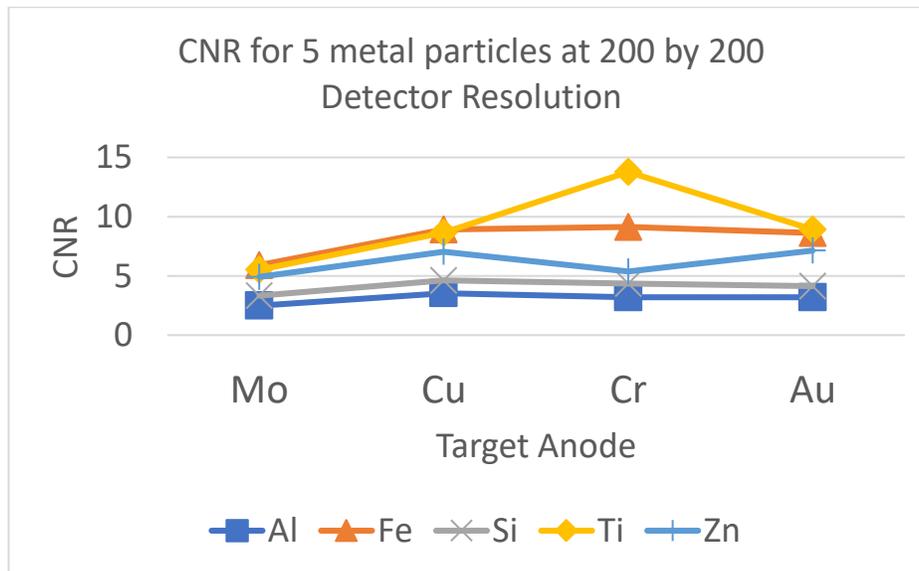

Fig. 8. Contrast to Noise Ratio, CNR, trendlines for attenuation images produced in MATLAB of the 5 metal particles of interest with 5 kV truncated spectra for each target anode with 0.25 micron detector pixel size

### 3.7 Limits of Detection

In this study, we examine the limits of detection for our proposed system using the reasonable parameters that resulted in the most promising results as we can see from the above simulations. These parameters are the 5 keV energy spectra for the Cr X-ray anode, the focal spot size of 4 μm, the detector pixel number of 100 by 100, and the detector pixel size of 0.5 μm. The simulation results are depicted in Fig. 9 for four different X-ray photon numbers: $10^9$ (Fig. 9a), $10^8$ (Fig. 9b), $10^7$ (Fig. 9c) and $10^6$ (Fig. 9d). We observed that the smaller samples would only appear if more x-rays were initialized. Furthermore, the 0.5 micrometer diameter targets are detectable when using $10^9$ and $10^8$ x-rays as shown in Fig. 9a and 9b. This performance of the simulated system is encouraging and so we plan to construct a similar physical system and test how it compares to the simulations.

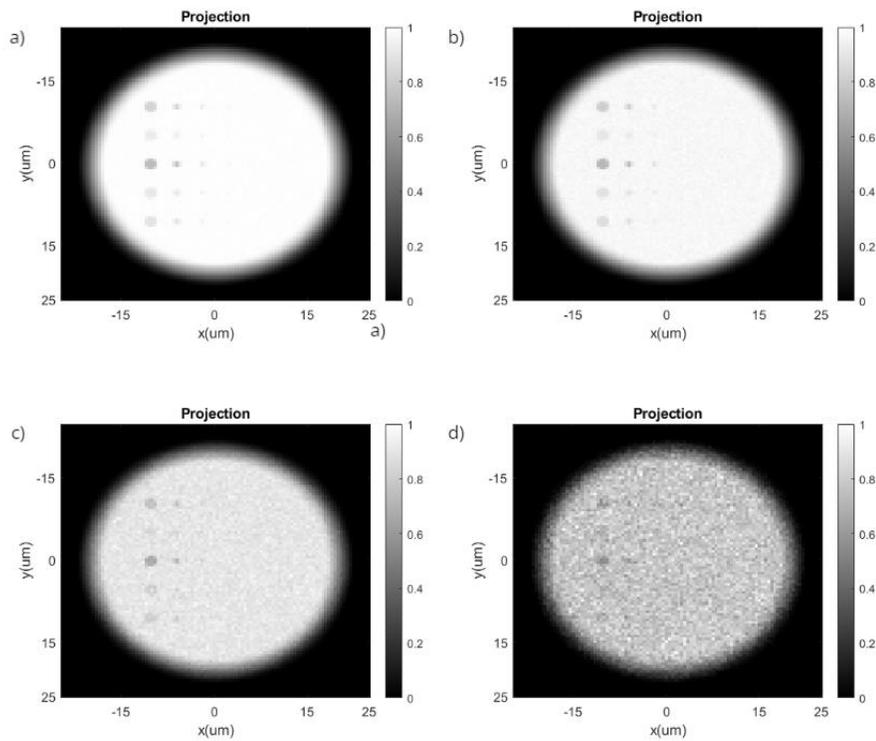

Fig. 9. Limit of detection test results. Each sample is arranged in a column with decreasing diameters going to the right. Projections captured with different X-ray photon number: (a) $10^9$; (b) $10^8$ ; (c) $10^7$; and (d) $10^6$. Cr source, 4 micrometers focal spot size, 100 by 100 pixel detector with 0.5 micron detector pixel length and 5 keV energy bandwidth were used for all.

### 3.8 Denoised images

**Table 6. Metal oxide calculated CNR for different x-ray photon counts after processing with Noise2Sim denoising algorithm**

| X-ray Count | CNR of Al (Standard deviation) | CNR of Fe | CNR of Si | CNR of Ti | CNR of Zn |
|---|---|---|---|---|---|
| $10^9$ | 22.2216 (0.3545) | 64.1866 (0.6344) | 30.7824 (0.6912) | 100.0272 (1.2846) | 37.6721 (0.4198) |
| $10^8$ | 9.7184 (0.2962) | 26.5366 (1.0076) | 12.6383 (0.4797) | 39.7463 (1.7712) | 15.5823 (0.7265) |
| $10^7$ | 2.7979 (0.6702) | 9.4348 (0.3064) | 4.1471 (0.1107) | 13.2797 (0.5635) | 5.1339 (0.2370) |
| $10^6$ | 0.9082 (0.1294) | 2.2196 (0.2063) | 1.4880 (0.2070) | 4.4389 (0.4697) | 1.8753 (0.0874) |

**4 micrometer focal spot size, 0.5 micrometer pixel size, and 5 keV energy bandwidth were used for all cases.**

The performance of the various denoising algorithms can be easily seen in Fig.10, which shows the boxplots obtained from the CNR values calculated from the images obtained after performing various denoising algorithms. From the boxplots, it can be clearly seen that BM3D provides the highest increase in the CNR values compared to the original noisy images. However, its CNR values also have the highest variance, and there are significant overlaps in the boxplots of Ti, Fe, Al, Si, and Zn for all doses. For $10^8$ and $10^9$ photon numbers, even the original noisy images have significant CNR separation, while BM3D denoised images do not provide that separation even if it increases the CNR values greatly. Thus, BM3D is practically infeasible for material identification using quantitative analysis. Noise2Sim, on the other hand, does not provide a significant increase in the CNR values, but it can provide the CNR separation by decreasing the variance, which is essential for material identification even for a low dose of $10^7$ photons. Thus, Noise2Sim is the most suitable algorithm for our purpose.

Here, the mean and standard deviations of the CNR values after denoising using the Noise2Sim method can be seen in Table 6. Comparing with Table 3, we can clearly see that the means of the CNR values have increased while the standard deviations of the CNR values have decreased for all photon numbers. We can see the effect of denoising on the projections in Fig. 11, where we can qualitatively see significant improvements in the noise condition even for low doses. Fig. 12 shows the image profiles taken across the two diagonals, where we can see that Noise2Sim can maintain the attenuation caused by the materials while decreasing the noise-related fluctuations everywhere else.

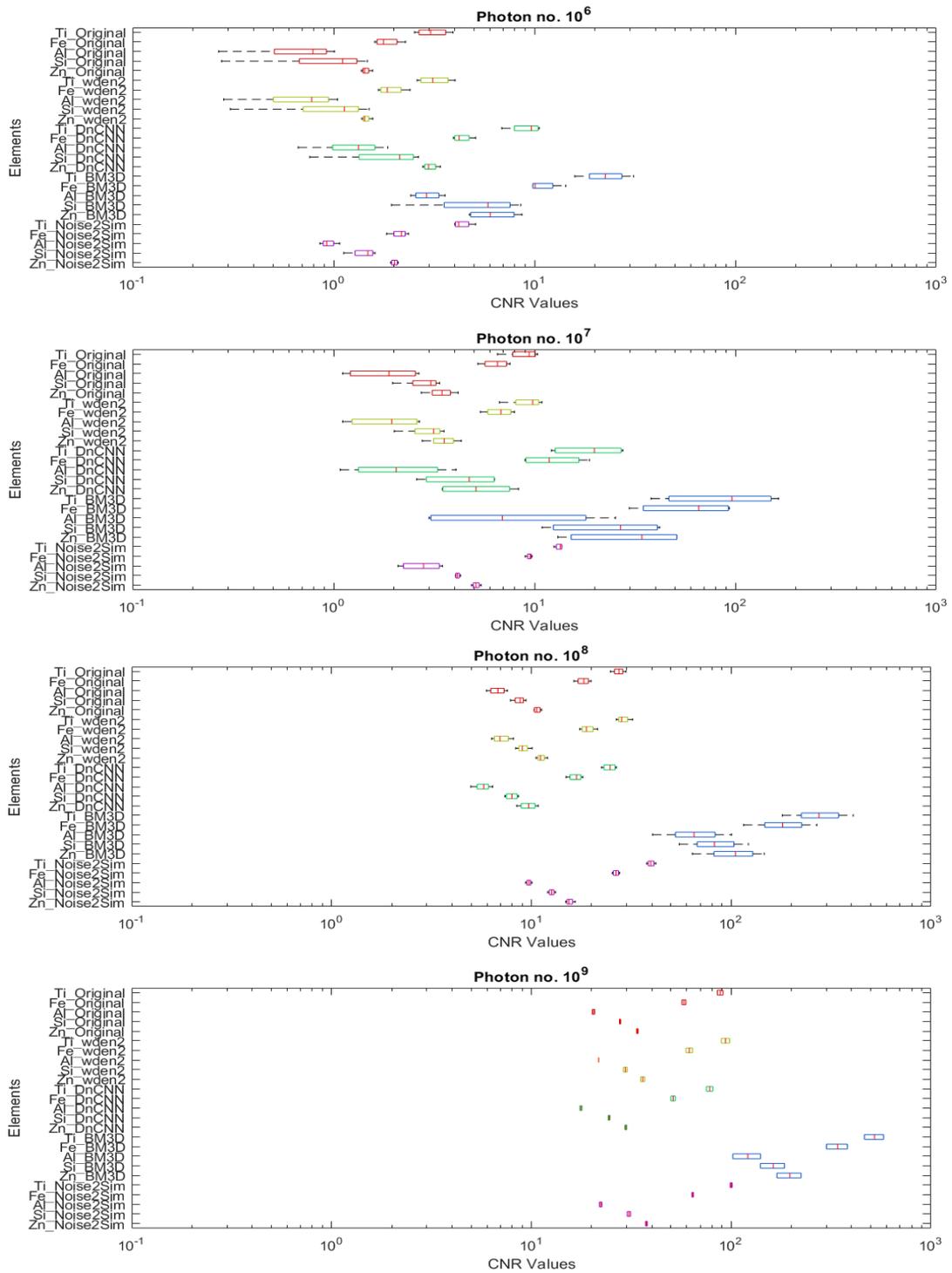

Fig. 10. Boxplot obtained from the CNR values calculated after performing various denoising algorithms (red—original projection, yellow—2D wavelet denoising, green—DnCNN, blue—BM3D, magenta—Noise2Sim). From the boxplots, the Noise2Sim method can provide sufficient separation even with a low dose of $10^7$ photons.

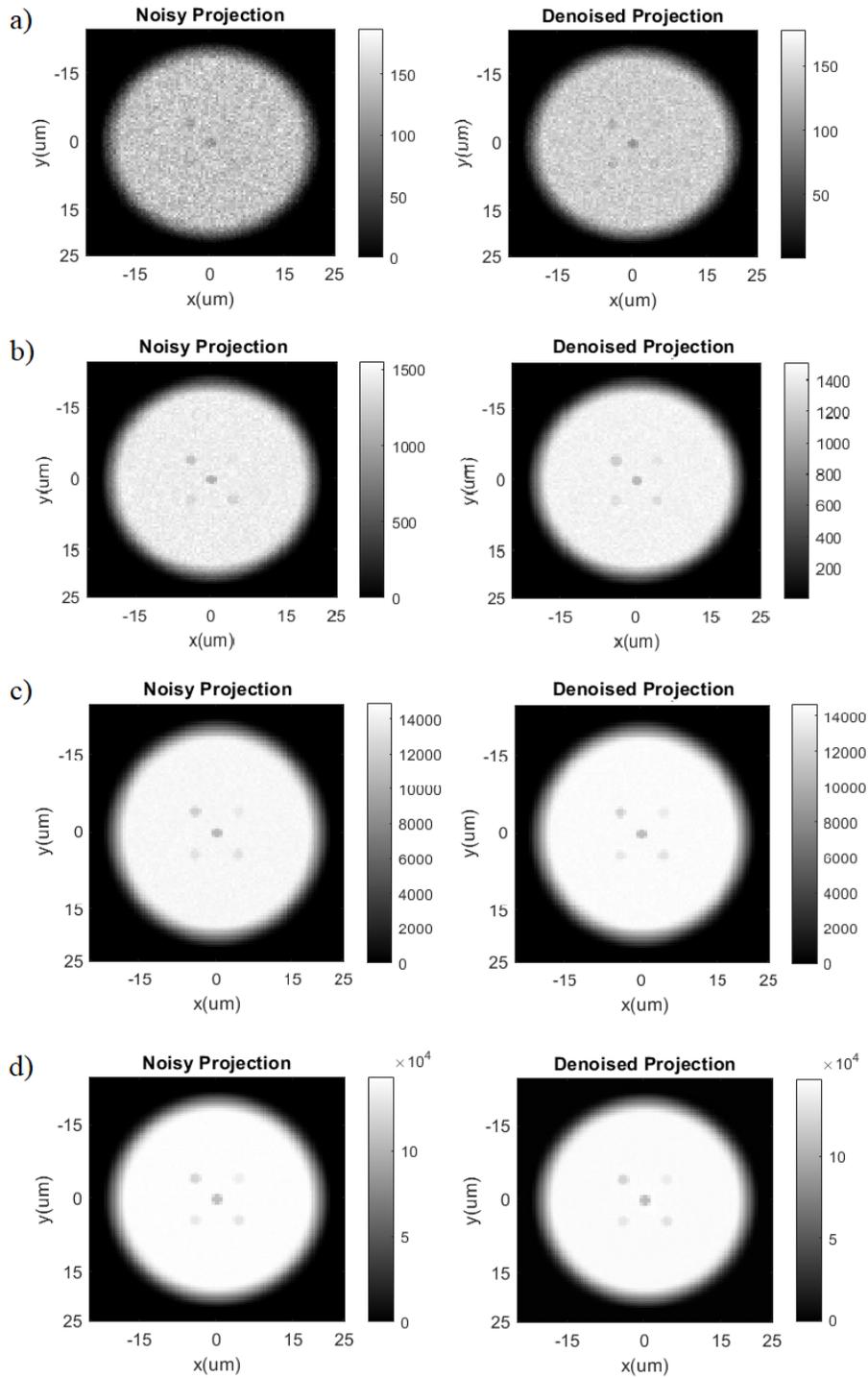

Fig. 11. Attenuation based Projection (not normalized) of the metal oxide sample using a 5 kV energy band spectra produced by the Cr target anode from the Sigray Source using (a). $10^6$, (b) $10^7$, (c). $10^8$, and (d). $10^9$ X-Rays before (left) and after (right) running through the Noise2Sim denoising algorithm.

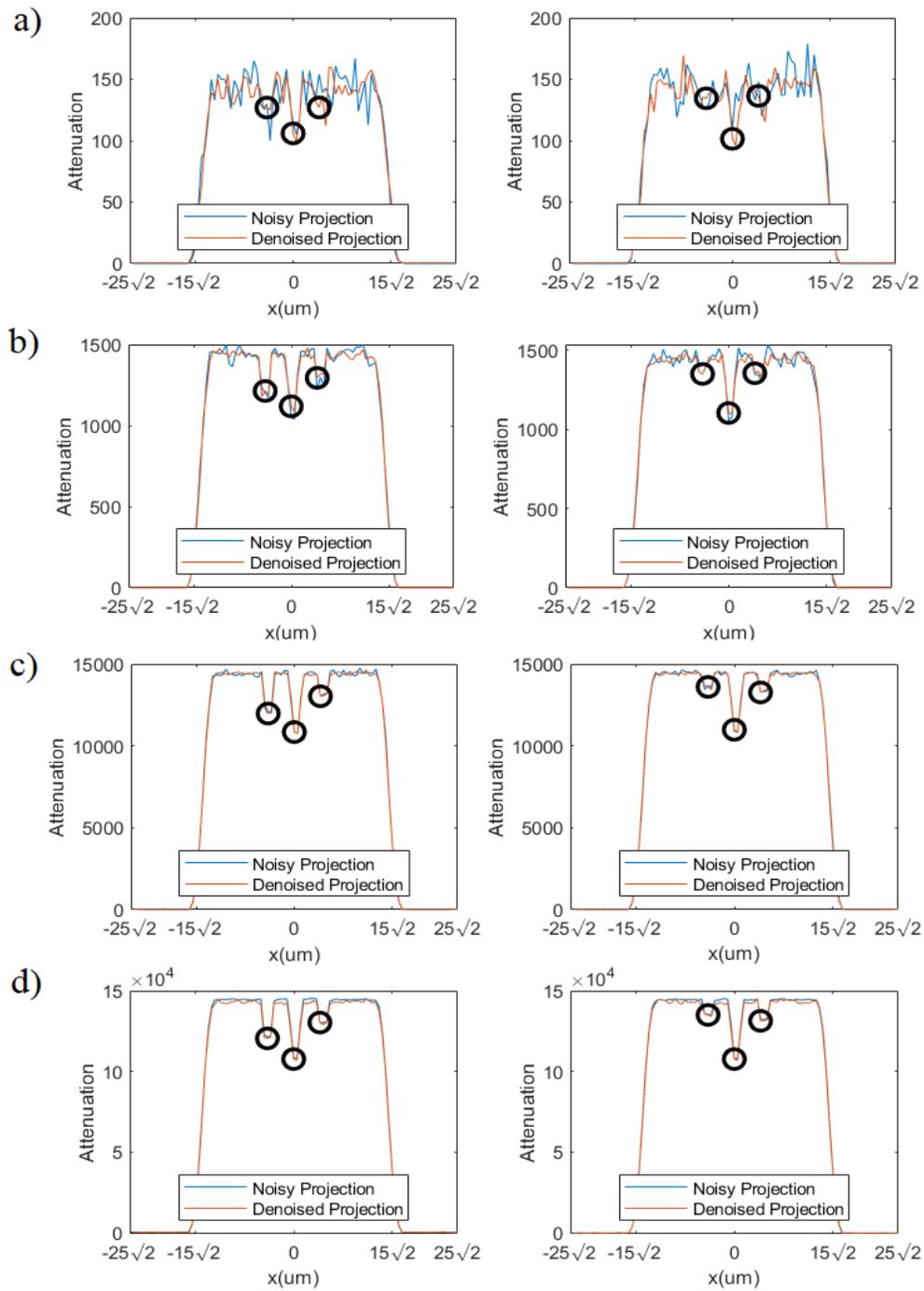

Fig.12. Image profile obtained across the major (left) and minor (right) diagonals of the projections obtained from (a). $10^6$, (b) $10^7$, (c). $10^8$, and (d). $10^9$ X-Rays before (blue) and after (red) running through the Noise2Sim denoising algorithm. From the profiles, especially (a) and (b), it is evident that the Noise2Sim model decreases the noise fluctuations while also maintaining the attenuation (dips) at the locations of the materials.

## 4. Discussion and conclusions

From the results of the above simulations, we have seen that these X-ray attenuation based images of the five different metal oxide particles inside gingival tissues are affected by many factors and our simulated results could guide the future imaging system design by analyzing the CNR changes due to the changes of these parameters. Due to the nature of X-ray attenuation, the metal oxide particles with lower densities such as the silicon dioxide and aluminum(III) Oxide do not have good contrast against the gingival tissue and the plastic plate background. Thus, it is critical for us to select narrow spectra bandwidth to enhance the CNR. As demonstrated by Fig. 4 and Table 2, we see that its CNR is best when we used the 5 keV spectra bandwidth for the X-ray metal anode of Cr.

Each of the four X-ray anode metals (Cr, Mo, Cu, and Au) has unique spectra that influenced the CNR of the five metal oxides targets inside gingival tissues. This is due to the differences in K-shell energies present in the spectra, and it was determined that truncating the spectra to 10 keV and below resulted in improvements in attenuation-based contrast. The K-shell energies of different target anodes such as Chromium, about 5.4 keV, are more abundant when the spectra are truncated from 10 kV and 5 keV.

Another systematic parameter that is important to the contrast of each metal oxide is the X-ray count. The higher X-ray photon counts result in more measurements to be recorded by the phase space detector in each pixel because it is very known that the signal noise ratio (SNR) is proportional to the square root of measured counts. Generally, the higher SNR results in better CNR. However, it is worth noting that higher X-ray photon number requires higher X-ray tube power and longer data acquisition time. To have the same CNR with fewer X-ray photon counts, we have applied denoising algorithms.

We also analyzed the effect of denoising on the X-ray projection and the material identification. To do so, we tested multiple denoising algorithms to see their performance. Wavelet-based image denoising is a popular and reliable denoising method that is being used by researchers for decades. Wavelet denoising, also known as wavelet thresholding, uses wavelet to decompose an image into various sub bands and then calculates the noise from those sub bands to remove them using hard or soft thresholding. It is a theoretically sound denoising method that has been used by researchers for decades. It is also used in many denoising-related articles as a standard benchmark of performance comparison.

DnCNN, proposed by Zhang et al [12], is a deep neural network that has been trained to perform blind Gaussian denoising with an unknown noise level as well as resolution enhancement. Even though its original training goal was to remove additive Gaussian noise, it has also been trained to perform image contrast and resolution enhancement. Thus, it can also effectively denoise images with non-Gaussian noise distributions. It is also popular among many researchers due to its easy-to-use MATLAB implementation using a pre-trained network. Since this work has mostly been implemented in MATLAB, it can be easily integrated with the DnCNN denoising algorithm. This is the main inspiration behind analyzing the quantitative performance of this particular method. Unfortunately, from Fig.10, it can be seen that while DnCNN improves the mean CNR value for $10^6$ and $10^7$ photon numbers, it does not provide the necessary separation for material identification. For higher doses, the mean CNR values tend to decrease instead of increase, even though there is an acceptable amount of CNR separation. For this reason, it is evident that the DnCNN denoising using a pre-trained model is not what we desire. It is also not possible to train this network using the scarce number of samples we have.

BM3D by Dabov et al [13] is another method that is widely popular and is also considered to be the state-of-the-art non-deep-learning-based denoising algorithm for general-purpose image denoising. Most of the recent denoising works, including BM3D and Noise2Sim, compare their performance to BM3D. BM3D at first creates a 3D array by stacking similar patches from an image frame, then exploit the correlation among the slices to calculate the noise and perform denoising. While both DnCNN and BM3D are very effective in qualitatively removing the noise and enhancing the signal-to-noise ratio for general images, in our case they not only perform over-smoothing on the data—thus increasing the mean CNR, but they also damage the quantitative information inside the X-ray projections, as can be seen from the boxplots in Fig. 10.

Our final algorithm, Noise2Sim by Niu et al [14], was selected mainly to overcome these issues. Noise2Sim has already been tested on X-ray-based CT and photon-counting CT scan images, and it is already shown that this method is capable of preserving the image characteristics while performing the denoising process. Noise2Sim is extremely compatible with our experiment since the experiment provides the requirements of the algorithm: multiple images with similar noise characteristics. The primary advantage of the Noise2Sim model thus in our case is the fact that it is a self-supervised model that only requires images with similar noise characteristics and no ground truth images for training. Being a self-supervised model, it needs far fewer samples compared to supervised models such as DnCNN. Thus, our dataset of 4 sample images and 3 no-sample images per photon number with no reference ground truth is suitable for the model.

Noise2Sim was trained separately for the different photon numbers with a learning rate of $10^{-4}$ for photon numbers $10^6$, $10^7$, and $10^8$; and $10^{-5}$ for photon number $10^9$. Adam optimizer was used to train the network with MAE as the loss function. Out of the 4 sample images, 2 were randomly selected as input and output to the network for training. Moreover, of the 3 no-sample images, 2 were randomly selected as input and output to the network for training. Either a pair of gum sample images (with metal oxides) or a pair of images without gum samples (only background noise) were used to train the network. A gum sample image and a no-sample image were never paired together to maintain the image similarity required for the Noise2Sim network.

Another systematic parameter investigated is whether the number of pixels in the detector allowed for an improved projection. The phase space actor makes this study much easier because the number of detector pixels could be produced after the simulation was completed and all the X-ray photon are recorded. This allows us to have much faster comparison of the effects that detector pixel size on the CNR for each of the five metal oxides. The effects proved to be interesting as each metal oxide has a unique change in their CNR when the pixel length is altered.

When we plot all the CNRs of five different metal oxides at different X-ray anodes (thus different X-ray spectra), we can see the significant differences of the changes in the CNR trends, from which we can differentiate five different metal oxide targets. It is worth noting that these CNR calculations have already included the effects of gingival tissues and the plastic plate. Thus, the simulations are close to the physical experiments. Advanced automatic diagnosis methods will be applied for the automatic differentiation of metal oxide particles in the future study because it is beyond the scope of this paper.

All the parameters investigated in this study show some levels of influence over the image quality. We have found that the system performs within the expected limits of detection with smaller metal oxide targets up to the size of each pixel being detected within a projection as indicated by Fig. 9. This encourages us to build the physical system to examine patient samples and determine whether they contain metal oxides which we have already examined in the simulation or others which can be implemented as well as tested.

In conclusion, our systematic simulations have investigated the X-ray attenuation based imaging and its capability in detection of metal oxide particles inside gingival tissues. We found that it is feasible to detect single metal oxide particles as small as 0.5 micrometers in diameter inside gingival tissue with a support plastic plate. We have also found that it is possible to differentiate these particles by using four different X-ray anode metals. Our findings are encouraging and will guide us to build a prototype X-ray projection imaging system to detect metal oxide particles and their clusters inside the actual gingival tissue from patient biopsies. We also plan to include a optical fluorescence imaging system sharing the field of view to study the correlations between the existing metal oxide particles and gingival tissue inflammation.

**Acknowledgement**

We would like to thank Sigray, Inc. for the spectral information of their Sigray x-ray source.

**Funding**

Work was done through the summer thanks to the Graduate Student Researcher Fellowship for Summer 2022 at the University of California, Merced.

This work was partially funded by the NIH National Institute of Biomedical Imaging and Bioengineering (NIBIB) [R01EB026646] and the National Institute of General Medical Sciences (NIGMS) [R42GM142394-01A1].